\pdfoutput=1
\documentclass[preprint,aps,pra,a4paper,american,floatfix,pdftex,superscriptaddress,citeautoscript
]{revtex4-1}%
\usepackage{amsfonts,amsmath,amssymb}
\usepackage[T1]{fontenc}
\usepackage{graphicx}%
\usepackage{textcomp}
\usepackage[utf8]{inputenc}
\usepackage{microtype}
\usepackage{titlesec}
\usepackage{xspace}
\usepackage{xcolor}
\usepackage{hyperref, hypernat}
\usepackage{mathtools}
\usepackage{caption}
\usepackage{subcaption}

\usepackage{chemformula}
\graphicspath{{./figures/}} 

\newcommand*{\um}{\ensuremath{\text{\textmu{m}}}\xspace}%
\newcommand*{\ie}{i.\,e.}%
\newcommand*{\uJ}{\ensuremath{\text{\textmu{J}}}\xspace}%

\newcommand{\ruimm}{\affiliation{Radboud University, Institute for Molecules
      and Materials, Heyendaalseweg 135, 6525 AJ Nijmegen, The Netherlands}}%
\newcommand{\mhmj}{\affiliation{MassSpecpecD BV, 7522LL Enschede, The Netherlands}}%

\begin{document}
\title{Photoelectron circular dichroism upon multiphoton ionization of a chiral alcohol}%
\author{Peter Krüger}\ruimm

\author{Michiel Balster} \ruimm
\author{Bhargava Ram Niraghatam} \ruimm
\author{Maurice H. M. Janssen} \mhmj

\author{\mbox{Daniel A.\ Horke}}%
\email[]{d.horke@science.ru.nl}%
\ruimm

\date{\today}%
\begin{abstract}\noindent
   We present the first photoelectron circular dichroism (PECD) measurements of chiral alcohols, and in particular 1-Phenylethanol, using multiphoton ionization at 400~nm. Observed PECD values were rather small at $\sim2$\%, but could be reliably extracted using both hemispherical integration and Abel inversion approaches. Experimental uncertainties of $<0.3$\% (2$\sigma$) where achieved with a collection time of around 2~hours. All experiments were conducted in a new compact spectrometer, featuring a continuous flow supersonic expansion and velocity-map imaging detection. The latter is crucial to extract reliable PECD values, as it allows discrimination of different features in the photoelectron spectrum, which exhibit different and opposing PECD signals. The use of a tabletop multiphoton universal ionization scheme is an important step towards a viable analytical chiral spectrometer based on PECD.
\end{abstract}
\maketitle

\section*{Introduction}
\label{sec:introduction}
The phenomenon of chiral molecules, that is molecules with non-superimposable mirror images, continues to fascinate scientists across a wide variety of disciplines. From the inherent chirality in biomolecules, leading to enantiomer-dependent bioactivities and the so-called \emph{homochirality of life}~\cite{Mason:Nature311:19,Ozturk:pnas119:e2204765119}, to fundamental physics and the quest for parity violation in chiral molecules~\cite{Tranter:Nature318:172,Fiechter:JPCL13:10011}.

Along with the general interest in chiral molecules ran the development of new chirally-sensitive spectroscopic techniques. While traditionally this was the domain of dichroic absorption spectroscopies, such as electronic or vibrational circular dichroism, several new approaches have emerged in the last decades, offering much higher sensitivities and chiral responses~\cite{Lehmann:JCP139:234307,Patterson:nature497:475,Ayuso:PCCP24:26962}. The concept of photoelectron circular dichroism (PECD) has been shown to be particularly suitable. This involves measuring the direction of outgoing photoelectrons following ionization with chiral (circularly polarized) light, with a PECD effect manifesting itself as an asymmetry in the forward and backward emission yield~\cite{Janssen:PCCP16:856}. PECD measurements have emerged as an attractive option for chiral analysis, since it frequently offers much higher chiral responses (sometimes 10s of \%~\cite{Ganjitabar:jms353:11}) than absorption-type spectroscopies, and can be combined with the high detection efficiency and throughput of charged-particle imaging~\cite{Chandler:JCP87:1445, Eppink:RSI68:3477}.

PECD was first theoretically described in the 1970s~\cite{Ritchie:PRA13:1411, Cherepkov:CPL87:344}, which was further refined in the early 2000s~\cite{Powis:JCP112:301, Powis:JPCA104:878}, before the first experimental observation some 30 years after the initial theoretical predictions. These first experiments utilized circularly polarized vacuum ultraviolet (VUV) synchrotron light sources~\cite{Bowering:PRL86:1187, Garcia:JCP119:8781}. Using single photon VUV photoionization for PECD detection has been a thriving field ever since~\cite{Nahon:JESRP204:322}, nowadays featuring dedicated endstations at synchrotron facilities~\cite{Garcia:RSI80:023102,Nahon:jsr19:508}. The realization of laboratory-based PECD measurements was shown around 10 years later~\cite{Lux:ACIE51:5001,Lehmann:JCP139:234307}, utilizing femtosecond resonance-enhanced multiphoton ionization (REMPI) and coincidence detection of the produced photoion and photoelectron~\cite{Vredenborg:RSI79:063108}. This was followed by the demonstration of the chiral analysis of complex mixtures using the same approach~\cite{Fanood:natcomm6:7511}. The laboratory-based PECD approach has also expanded significantly in the past years~\cite{Janssen:PCCP24:24611}, and recent technological demonstrations include the use of nanosecond high-resolution REMPI as an ionization source~\cite{Kastner:PCCP22:7404,Ranecky:PCCP24:2758}, high-harmonic based XUV light sources~\cite{Ferre:natphoton9:93}, the incorporation of novel molecular sources~\cite{Sparling:PCCP25:6009}, and high-throughput experiments with fiber-based high repetition-rate femtosecond lasers~\cite{Comby:natcomm9:5212,Comby:chir32:1225}. Chiral analysis using PECD has also been demonstrated for photodetachment from gas-phase anions recently~\cite{Kruger:acie60:17861, Triptow:ACIE62:e202212020}, including the first chirality measurement of larger biopolymers~\cite{Kruger:jpcl13:6110}.

\begin{figure}
    \centering
    \includegraphics[width=0.4\textwidth]{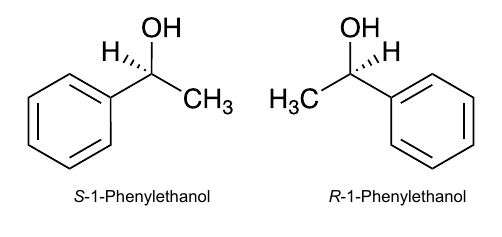}
    \caption{Structures of 1-Phenylethanol enantiomers.}
    \label{fig:structures}
\end{figure}

In this contribution we report the first PECD measurements of the two enantiomers of 1-Phenylethanol, as shown in \autoref{fig:structures}. These were performed using femtosecond multiphoton ionization (fs-MPI) at 400~nm, and hence demonstrating chiral analysis with a fully universal table-top ionization approach. The resulting PECD was rather small, $\sim$2\%, but could be measured with an absolute error of <0.3\% in a 2 hour long measurement with a 3~kHz repetition rate laser system. These measurements were conducted on a new and compact PECD spectrometer, specifically designed to be cost and space effective, while still making use of the benefits of a continuous cold supersonically-expanded molecular beam, and velocity-map imaging photoelectron detection~\cite{Eppink:RSI68:3477}. Since the spectrometer makes use of a continuous molecular beam, the duty cycle is limited only by the repetition rate of the laser and hence could be improved by 2--3 orders of magnitude by the use of high repetition-rate fiber-based femtosecond light sources.

\section*{Experimental Methods}
\label{sec:exp-methods}
\begin{figure*}
	\includegraphics[width=0.7\textwidth]{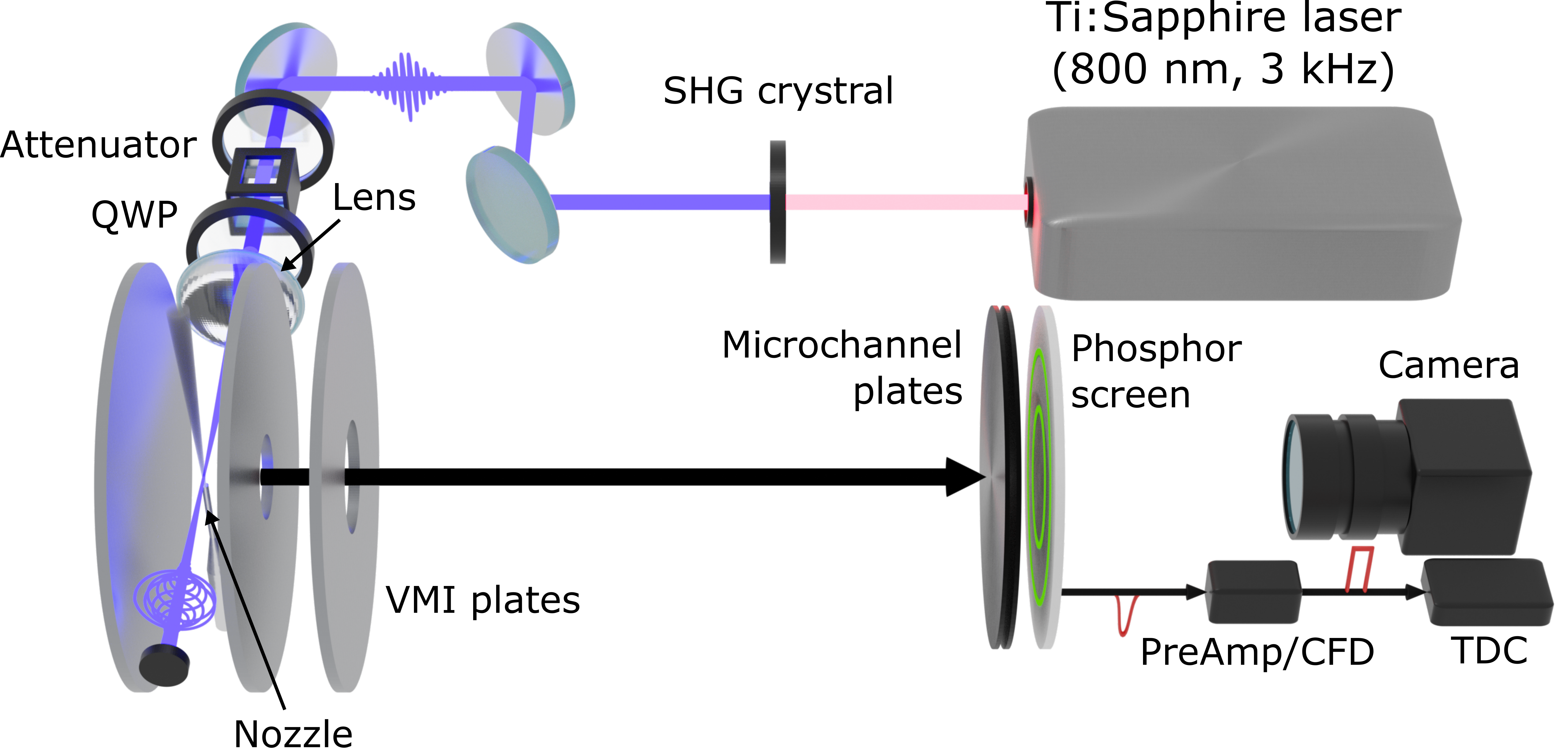}
	\caption{Schematic of the experimental setup. Molecules are supersonically expanded into a vacuum through a 15~\um orifice capillary, located 5~mm from the interaction point at the center of a velocity-map imaging (VMI) spectrometer. Molecules are ionized by circularly polarized 100~fs 400~nm laser pulses, with the helicity controlled by a motorized quarter-wave plate. Electrons or ions are accelerated by the VMI electrodes towards a position sensitive detector, \emph{via} a $\mu$-metal shielded field-free flight tube. For mass spectra the detector response is picked up by a constant-fraction discriminator and time stamped by a time-to-digital converter. Photoelectron images  are recorded with a fast CMOS camera and centroided on-the-fly.}
	\label{fig:setup}
\end{figure*}
Experiments were conducted in a newly designed compact velocity-map imaging (VMI) spectrometer, a schematic of the new experimental setup is shown in \autoref{fig:setup}. This consists of two vacuum chambers, with differential pumping provided by the VMI electrodes that separate the chambers. The compact source and interaction chamber is based on a CF63 6-way cross and houses the VMI electrodes and gas nozzle. It is pumped by a small turbomolecular pump (HiPace 300, Pfeiffer Vacuum), with a typical operating pressure (under load) of $1\times 10^{-5}$~mbar. The detection chamber provides a $\mu$-metal shielded field-free flight tube for electrons or ions of 400~mm, before particles are detected on microchannel plate (MCP) detector, coupled to a fast phosphor screen (Photonis ADP 2PS, 40 mm, 1:60, P47 phosphor). This chamber is pumped by a 700 l/s pump (HiPace 700, Pfeiffer Vacuum), with typical operating pressures under load of $<1\times 10^{-6}$~mbar.

Sample molecules are introduced into the spectrometer by means of supersonic expansion from a handmade glass nozzle. The nozzle tip has an opening diameter of about 15~$\mu$m and is coated in silver to allow high voltages to be applied. The nozzle is placed equidistant between the extractor and repeller electrode of the VMI setup. A high voltage is applied to the silver coated nozzle tip to minimize distortions of the electric field due to the tip, and it is typically operated at the average voltage of repeller and extractor (\ie, $V_\text{tip}=\frac{1}{2}(V_\text{rep}+V_\text{extr})$). The tip of the nozzle is located $\sim5$~mm from the interaction point in the center of the VMI, where the expanding gas plume is intersected orthogonally by a focussed laser beam. The VMI is based on the classic 3-plate design from Eppink \& Parker~\cite{Eppink:RSI68:3477}, with an outer diameter of 40~mm, and central orifices of 5~mm (extractor) and 10~mm (ground).

Produced charged particles are accelerated towards the detector, where we either collect time-of-flight mass spectra or photoelectron images. For the former, the response from the MCP is counted after pre-amplification and constant-fraction-discrimination (Surface Concept 1-Channel Preamplifier-CFD) using a high-resolution time-to-digital card (chronologic xTDC4-PCIe, 16~ps bins). For electron detection, impact positions on the phosphor screen are recorded with a fast CMOS camera (Basler acA720-520um), operating at a frame rate of around 1~kHz and a resolution of $256\times256$ pixel. To increase resolution and overcome any spatial inhomogeneities of the detector response, events are centroided on-the-fly and only centroids retained for further analysis. 

As a light source a 3~kHz Ti:Sapphire laser (Spectra Physics SpitfireAce) was used, producing fundamental output centered at 800~nm and with 100~fs pulse duration. Photons at 400~nm were generated by frequency doubling in a beta-barium borate (BBO) crystal. The pulse energy was controlled by an attenuator consisting of a half-wave plate (CASIX WPZ1315) and a linear Glan laser polarizer.  Circular polarized light was then generated by an achromatic quarter-wave plate (B. Halle 300-470~nm). The produced helicity could be inverted by rotating the waveplate 90 degrees using a motorized rotation stage (Standa 8MRU-1). Both circular polarization states were characterized by rotating a linear polarizer (Thorlabs GL10B) behind the quarterwave-plate while measuring the transmitted power. Stokes vectors of (1, 0.06, 0.034, -0.998) and (1, 0.053, 0.039, 0.998) were determined for LCP and RCP, respectively (see supplementary information for further details). The laser is focussed into the VMI spectrometer using a plano-spherical lens (Thorlabs LA4904-UV, f = 15~cm), yielding typical spotsizes of 15~\um (FWHM).  

A single PECD measurement consisted of 30000 camera frames for each of the two polarizations, and hence takes about 1~min. For the data presented here, the mean PECD and corresponding standard errors were determined from a total of 260 measurements, confidence intervals are given as two standard errors. During data acquisition the polarization state sequence is alternated to prevent systematic errors by long-term drifts, i.e. data is collected in the sequence (LCP-RCP)-(RCP-LCP)-....  Photoelectron images were analyzed using Python scripts including the pyAbel package~\cite{Hickstein:pyabel}, utilizing the rBasex method for inversion with finite differences regularization (Strength = 30000)~\cite{rBasex:0.9.0}.

Samples of (S)- and (R)-Phenylethanol, each with a nominal purity of 97\%, were purchased from Merck Life Science and used without further purification. The sample vapour (at room temperature) is picked up and transported to the nozzle by a small Helium flow (300~mbar) for expansion into the vacuum chamber of the spectrometer.

\section*{Results and Discussion}
\label{sec:results}
\begin{figure*}
	\includegraphics[width=\textwidth]{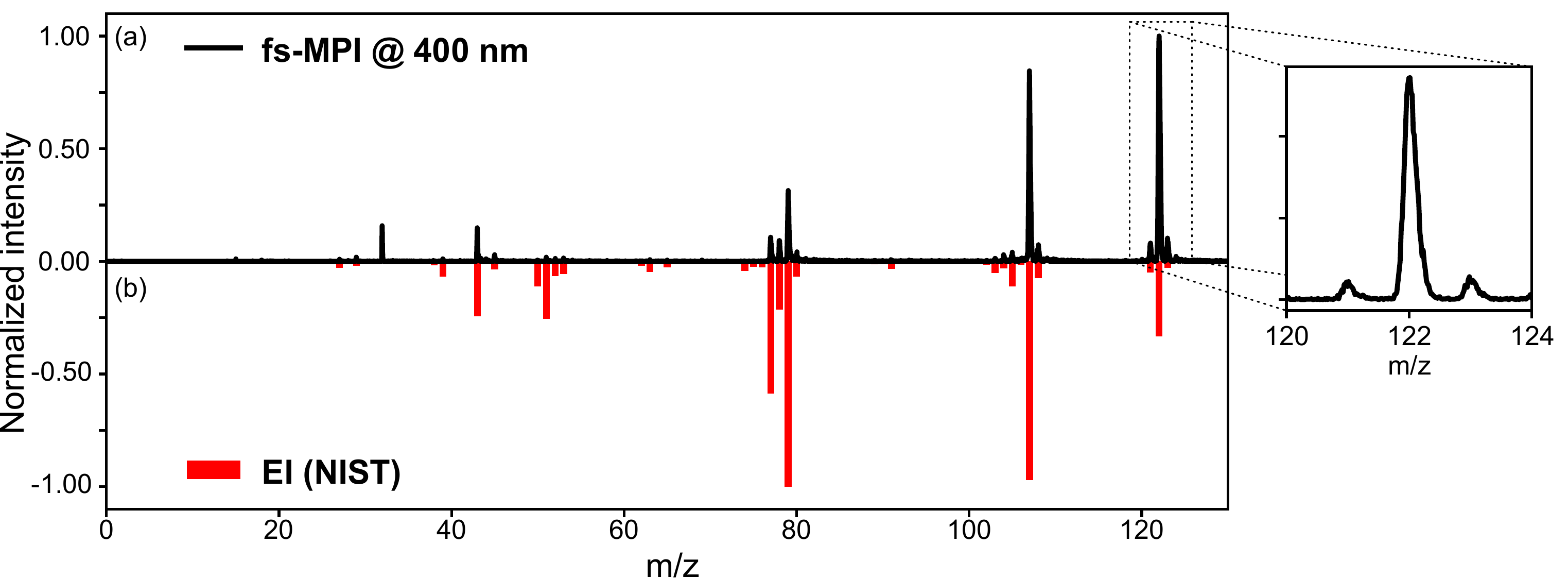}
	\caption{Mass spectra of R-1-Phenylethanol. (a) Time-of-flight mass spectrum acquired in our new spectrometer using femtosecond multiphoton ionization at 400~nm. The inset shows in detail the mass region of the parent ion, indicating the achievable mass resolution of our spectrometer. (b) Reference electron impact ionization mass spectrum from the NIST database~\cite{NIST:webbook:MassSpec}.}
	\label{fig:mass_spec}
\end{figure*}

An exemplary 400~nm multiphoton ionization mass spectrum of R-1-Phenylethanol is shown in the top half of \autoref{fig:mass_spec}. The inset shows a detailed view of the m/z region $120-124$, which yields an experimental mass resolution (FWHM) of $\frac{M}{\Delta M}=523$. The mass spectrum is dominated by the parent ion at 122~Da, with main fragments appearing at m/z = 43, 77, 78, 79, and 107. For comparison, we also show a literature (NIST chemistry webbook) electron impact ionization mass spectrum in the lower half of \autoref{fig:mass_spec}~\cite{NIST:webbook:MassSpec}. Its is clear that femtosecond multiphoton ionization (fs-MPI) is a relatively soft process, inducing significantly less fragmentation than electron impact ionization, and a much larger fraction of intact parent ions~\cite{Dauletyarov:JASMS34:1538}. Apart from a small contamination appearing at m/z = 32, similar fragments are encountered in both spectra, allowing the clear identification of 1-Phenylethanol.  

\begin{figure}
	\includegraphics[width=0.5\textwidth]{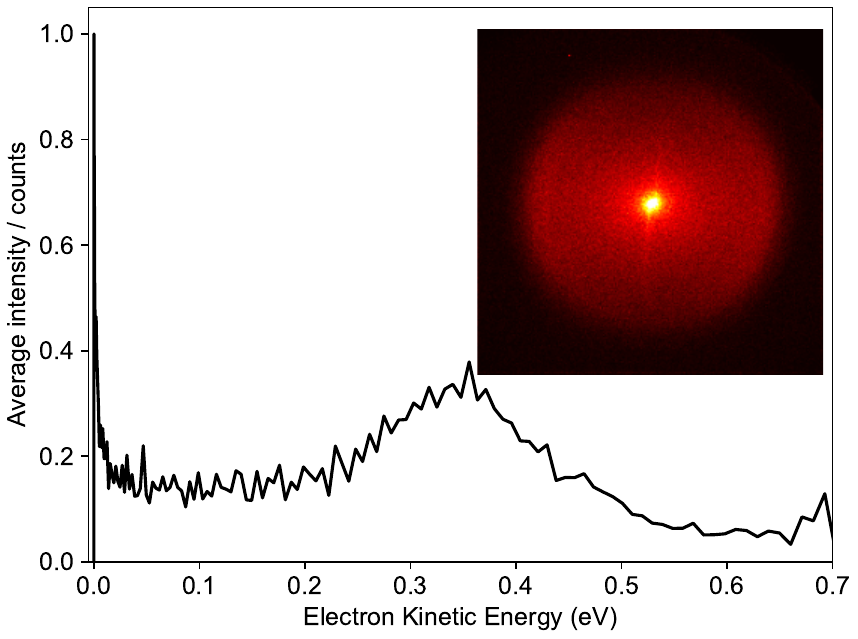}
	\caption{Photoelectron spectrum and raw photoelectron image (inset) of 1-Phenylethanol at reduced laser power to avoid saturation effects in the center of the image. }
	\label{fig:pes_lowcounts}
\end{figure}

We now consider the photoelectron image and spectrum of 1-PE collected at 400~nm and shown in \autoref{fig:pes_lowcounts}, collected at laser pulse energies of 1.5~\uJ (corresponding to $
\sim2\times10^{12}$ W/cm$^2$). The photoelectron images exhibited two features; a sharp and intense feature at very low electron kinetic energy (eKE) in the center of the image, and a weaker outer ring at higher eKE. The latter feature peaks at around 0.35~eV electron kinetic energy. The (non-adiabatic) ionization energy for 1-PE has been reported as 8.89~eV~\cite{Mons:PCCP2:5065}, ionization at 400~nm thus requires 3 photons which carry a total photon energy of 9.3~eV. Assuming such a 3-photon process, our data yields a vertical (non-adiabatic) ionization energy of $8.95\pm 0.1$~eV, fully consistent with the literature value. The observed feature is significantly broader than our experimental resolution and bandwidth (see S.I. for details), most likely indicating a significant change in molecular geometry upon photoionization, as has been noted before~\cite{Mons:PCCP2:5065}. From the high-energy cutoff at around 0.6~eV in \autoref{fig:pes_lowcounts} we can estimate the adiabatic ionization energy as around 8.7~eV.

To obtain enantiosensitive information, photoelectron images were obtained for both helicities of circularly polarized light. For these measurements we focus only on the weaker outer feature, since a feature in the center of the image does not yield information on the photoelectron angular distribution and hence photoemission asymmetry. The used laser pulse energy was hence increased to 2~\uJ (fluence of $\sim3\times10^{12}$ W/cm$^2$) to obtain sufficient count rates in the high eKE region. This caused centroiding artifacts in the image center due to overlapping events and this region has therefore been masked out in the images shown here.

\begin{figure*}
    \includegraphics[width=0.8\columnwidth]{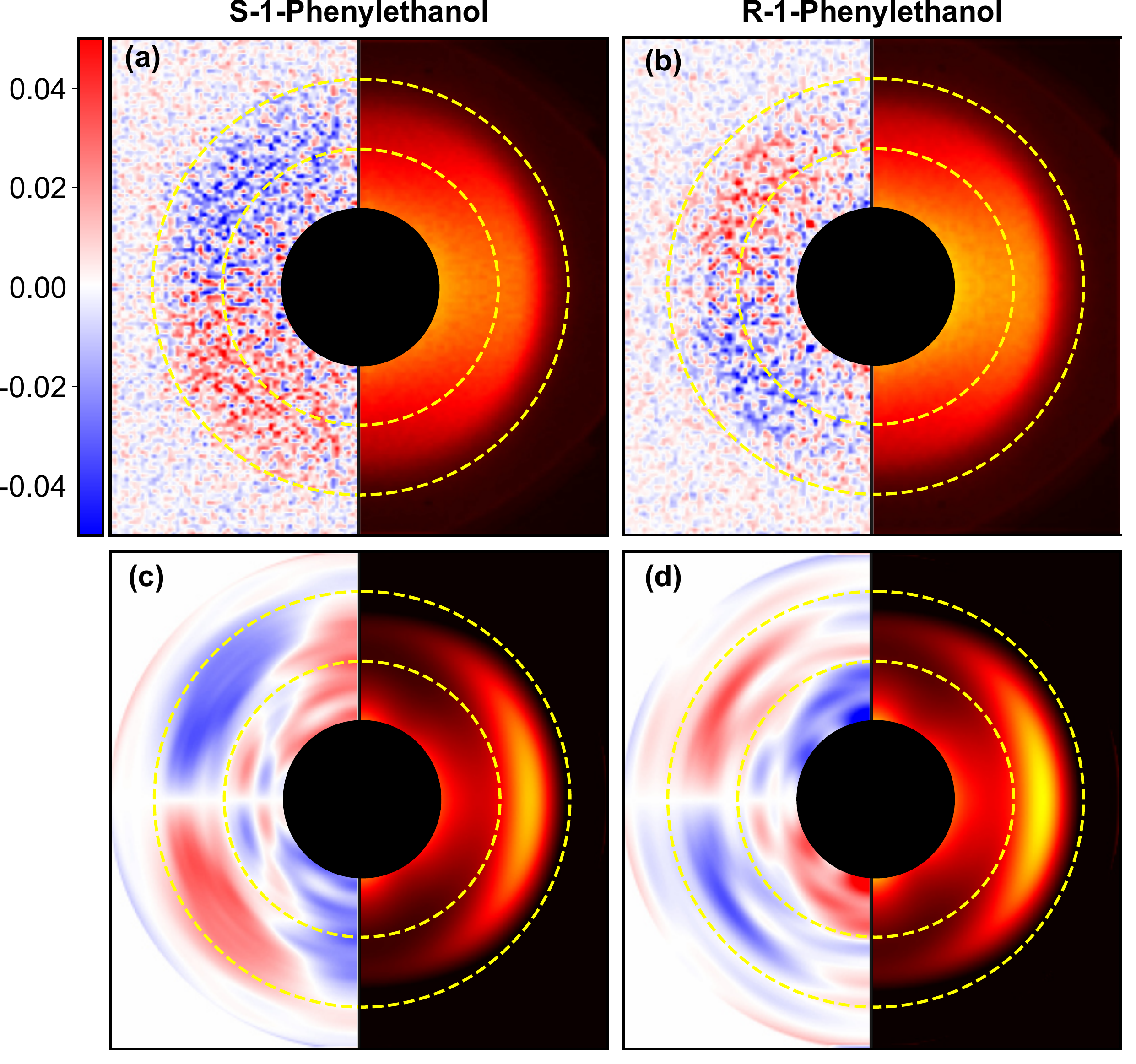}
     \caption{Antisymmetric difference (left) and symmetric sum (right) photoelectron images based on raw (a, b) as well as Abel-inverted data (c, d) for S- (a, c) and R-1PE (b, d), respectively. For enhanced contrast the raw images where $2\times$ binned, resulting in $128\times128$ pixel images. The central region was masked due to artifacts caused by abundant low kinetic energy electrons. Dashed circles indicate the radial region used for  calculation of the PECD effect, see text for details. The laser propagation direction is upwards in all images.}
     \label{fig:images}
\end{figure*}

Photoelectron images for both enantiomers and both helicities are shown in \autoref{fig:images}, where the left half always corresponds to the antisymmetric difference image, and the right half to the symmetric sum image of the two enantiomers. The top panels (a,b) show raw images, whereas the bottom panels (c,d) show Abel-inverted images. The center of all images has been masked out, as explained above.

The anti-symmetrized difference images (LCP-RCP) for R- and S-1PE show a clear forward-backward asymmetry along the laser propagation axis, with photoelectrons for R-1PE preferentially emitted in the forward direction for LCP light, and correspondingly in the backward direction for S-1PE. This is apparent both in the raw difference images as well as in the Abel-inverted images. This PECD effect can be quantized in a single value by either simple hemispherical integration or based on the Legendre coefficients of the photoelectron angular distributions. For this only the radial region of the images containing the high-eKE features was used, as indicated by the dashed yellow circles in \autoref{fig:images}. The range used was 0.21--0.49 eV, corresponding to the FWHM of the high eKE feature.
For the hemispherical integration approach, the PECD is then calculated from the respective electron yields in the forward (\emph{fwd}) and backward (\emph{bwd}) direction for both helicities of light (\emph{rcp,lcp})~\cite{Janssen:PCCP16:856,Lehmann:JCP139:234307}:
\begin{equation}
    \mathrm{PECD} = 2 \left( \frac{Y_{lcp}^{fwd} - Y_{lcp}^{bwd}}{Y_{lcp}^{fwd} + Y_{lcp}^{bwd}} - \frac{Y_{rcp}^{fwd} - Y_{rcp}^{bwd}}{Y_{rcp}^{fwd} + Y_{rcp}^{bwd}} \right).
    \label{eq:hemi-pecd}
\end{equation}
This approach yielded PECD values of $+1.67 \% \pm 0.26 \%$ and $-2.25 \% \pm 0.26 \%$ for R- and S-1PE, respectively. Throughout, all given uncertainties correspond to 2 standard errors.

Alternatively, we can define the PECD based on the (odd) Legendre coefficients ($\beta_n$) contributing to the Abel-inverted photoelectron image~\cite{Cooper:JCP48:942}. For a 3-photon process Legendre polynomials up to order $n=5$ need to  be taken into account, and hence we define the PECD as~\cite{Lehmann:JCP139:234307}:
\begin{equation}
    \mathrm{PECD} = 2 \beta_1 - \frac{1}{2} \beta_3 + \frac{1}{4} \beta_5.
    \label{eq:abel-pecd}
\end{equation}
This yielded PECD values of $+1.63 \% \pm 0.30 \%$ and $-2.08 \% \pm 0.28 \%$ for R- and S-1PE, respectively.
\begin{table}
    \centering
    \caption{PECD values of enantiopure 1-PE samples determined by hemispherical integration of raw images and PAD fitting of Abel-inverted images. Uncertainties are given as $2\sigma$.}
    \label{tab:pecd}
    \begin{tabular}{ l l l }
    \hline
    Sample \hspace{2mm} & PECD (Abel inversion) \hspace{2mm}& PECD (hemisph. int.) \\ \hline
    R-1PE & $+1.63 \pm 0.30$ & $+1.67 \pm 0.26$  \\  
    S-1PE & $-2.08 \pm 0.28$ & $-2.25 \pm 0.26$ \\ \hline
    \end{tabular}
\end{table}

The corresponding values for the two enantiomers are also summarized in \autoref{tab:pecd}. PECD values extracted from both analysis frameworks are in excellent agreement and show the expected sign inversion. They furthermore agree with a recent PEELD study on chiral alcohols~\cite{Greenwood:PCCP25:16238}. The S-1PE enantiomer consistently exhibited a larger PECD effect then measured for the R-1PE, potentially indicating that our sample of R-1PE had a reduced enantiopurity. This direct comparison of the two analysis pathways shows that for reliable quantitative PECD determination a full Abel analysis of the VMI images is not required (at least as long as there are no strongly overlapping features), and the simpler hemispherical integration is sufficient. Indeed the slightly larger uncertainties for Abel inversion are likely due to the reduced number of electrons taken into account when only considering the central slice of the distribution. The use of the VMI technique, and hence the ability to define electron kinetic energy ranges over which to determine the PECD, is highly beneficial compared to simple integrating detectors such as half-moon anodes or electron multipliers. It allows differentiation of different features in the photoelectron spectrum that frequently correspond to different electronic states in the cation, with potentially very different PECD effects that cannot be distinguished in simple integrating detectors. Indeed, this is also the case for 1PE here. Abel inverted images in \autoref{fig:images} clearly show a feature with inverted PECD asymmetry at lower radii, albeit with much reduced intensity. Nonetheless, a simple integrating detector would have recorded much lower PECD values of 1.00 (R-1PE) and -1.42 (S-1PE).

Abel analysis furthermore lets us analyze the contributions from individual Legendre coefficients to the overall observed PAD. The PECD effect is contained only in the odd Legendre coefficients, and the extracted values for the two enantiomers are shown in \autoref{fig:betas}. Numerical values for all extracted coefficients are given in the supplementary information. This clearly shows that the main contributors to the PECD are the $\beta_1$ and $\beta_3$ terms, which also show the expected mirroring behavior.
\begin{figure}
    \centering
    \includegraphics[width=0.5\textwidth]{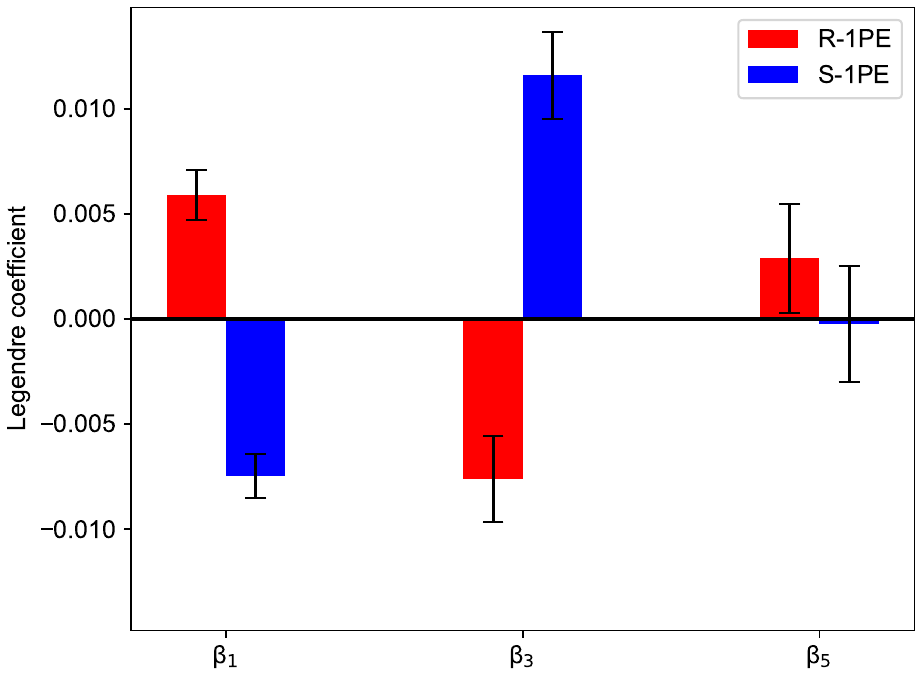}
    \caption{Amplitudes of the odd Legendre coefficients for both enantiomers extracted from the PAD fitting. The main contribution to the PECD comes from the $\beta_1$ and $\beta_3$ terms, with all orders showing the expected sign inversion between the enantiomers.}
    \label{fig:betas}
\end{figure}

While the magnitude of the PECD effect of only $2~\%$ is small compared to many previously investigated substances, it is clearly measurable with high fidelity in our compact spectrometer. The extracted values are furthermore comparable with a recently published photoelectron elliptical dichroism (PEELD) study, which utilized (3+1) multiphoton ionization at 520~nm to study (amongst others) chiral alcohols~\cite{Greenwood:PCCP25:16238}. A previous study already demonstrated the recording of PECD signals in the region of above-threshold ionization (ATI) and tunnel ionization~\cite{Beaulieu:njp18:102002}, with the current study we are within the multiphoton ionization region (Keldysh parameter $\gamma\sim 2$). This confirms the earlier observation that even with strong-field ionization, whether it is ATI, tunneling or MPI, the PECD effect still persists.

\section*{Conclusion}
\label{sec:conclusion}
We have shown the first PECD measurements of 1-Phenylethanol using femtosecond multiphoton ionization. This highlights the universality of the fs-MPI approach for chiral analysis. The PECD for 1-PE, albeit small at only $\sim2$\%, was reliably extracted using both hemispherical integration and Abel analysis, and showed the expect mirroring behavior. Experiments were conducted on a new compact PECD spectrometer, featuring a small diameter nozzle and optimized pumping geometry, which enables a small footprint mobile machine, with much reduced pumping requirements compared to conventional gas-phase photoelectron spectrometers. The use of VMI detection yields additional spectroscopic information on the system and allows the filtering out of particular photoelectron features for reliable PECD determination. In the current iteration using a 3~kHz laser, reliable quantitative PECD analysis can be performed in around one hour. We are currently working on implementing a high repetition-rate femtosecond laser that will improve the duty cycle, and hence reduce the measurement time, by 2--3 orders of magnitude. This will pave the way for integration of PECD-based chiral analysis with standard analytical separation techniques such as chromatography.

\section*{Acknowledgements}
This work was supported by the Netherlands Organization for Scientific Research (NWO) under grant numbers STU.019.009, VIDI.193.037 and 712.018.004, and the European Regional Development fund (EFRO, OP Oost) under project number PROJ-00949. We furthermore thank the Spectroscopy of Cold Molecules Department, and in particular Prof. Bas van de Meerakker, for continued support.

\section*{Conflicts of Interest}
MHMJ is founder and CEO of \emph{MassSpecpecD BV} (www.MassSpecpecD.com).

\clearpage
\bibliography{string,UCD-bib}

\providecommand{\latin}[1]{#1}
\makeatletter
\providecommand{\doi}
  {\begingroup\let\do\@makeother\dospecials
  \catcode`\{=1 \catcode`\}=2 \doi@aux}
\providecommand{\doi@aux}[1]{\endgroup\texttt{#1}}
\makeatother
\providecommand*\mcitethebibliography{\thebibliography}
\csname @ifundefined\endcsname{endmcitethebibliography}
  {\let\endmcitethebibliography\endthebibliography}{}
\begin{mcitethebibliography}{41}
\providecommand*\natexlab[1]{#1}
\providecommand*\mciteSetBstSublistMode[1]{}
\providecommand*\mciteSetBstMaxWidthForm[2]{}
\providecommand*\mciteBstWouldAddEndPuncttrue
  {\def\EndOfBibitem{\unskip.}}
\providecommand*\mciteBstWouldAddEndPunctfalse
  {\let\EndOfBibitem\relax}
\providecommand*\mciteSetBstMidEndSepPunct[3]{}
\providecommand*\mciteSetBstSublistLabelBeginEnd[3]{}
\providecommand*\EndOfBibitem{}
\mciteSetBstSublistMode{f}
\mciteSetBstMaxWidthForm{subitem}{(\alph{mcitesubitemcount})}
\mciteSetBstSublistLabelBeginEnd
  {\mcitemaxwidthsubitemform\space}
  {\relax}
  {\relax}

\bibitem[Mason(1984)]{Mason:Nature311:19}
Mason,~S.~F. Origins of Biomolecular Handedness. \emph{Nature} \textbf{1984},
  \emph{311}, 19--23\relax
\mciteBstWouldAddEndPuncttrue
\mciteSetBstMidEndSepPunct{\mcitedefaultmidpunct}
{\mcitedefaultendpunct}{\mcitedefaultseppunct}\relax
\EndOfBibitem
\bibitem[Ozturk and Sasselov(2022)Ozturk, and
  Sasselov]{Ozturk:pnas119:e2204765119}
Ozturk,~S.~F.; Sasselov,~D.~D. On the origins of life’s homochirality:
  Inducing enantiomeric excess with spin-polarized electrons. \emph{Proc.\
  Natl.\ Acad.\ Sci.\ U.S.A.} \textbf{2022}, \emph{119}, e2204765119\relax
\mciteBstWouldAddEndPuncttrue
\mciteSetBstMidEndSepPunct{\mcitedefaultmidpunct}
{\mcitedefaultendpunct}{\mcitedefaultseppunct}\relax
\EndOfBibitem
\bibitem[Tranter(1985)]{Tranter:Nature318:172}
Tranter,~G.~E. Parity-Violating Energy Differences of Chiral Minerals and the
  Origin of Biomolecular Homochirality. \emph{Nature} \textbf{1985},
  \emph{318}, 172--173\relax
\mciteBstWouldAddEndPuncttrue
\mciteSetBstMidEndSepPunct{\mcitedefaultmidpunct}
{\mcitedefaultendpunct}{\mcitedefaultseppunct}\relax
\EndOfBibitem
\bibitem[Fiechter \latin{et~al.}(2022)Fiechter, Haase, Saleh, Soulard,
  Tremblay, Havenith, Timmermans, Schwerdtfeger, Crassous, Darquié, Pašteka,
  and Borschevsky]{Fiechter:JPCL13:10011}
Fiechter,~M.~R.; Haase,~P. A.~B.; Saleh,~N.; Soulard,~P.; Tremblay,~B.;
  Havenith,~R. W.~A.; Timmermans,~R. G.~E.; Schwerdtfeger,~P.; Crassous,~J.;
  Darquié,~B.; Pašteka,~L.~F.; Borschevsky,~A. Toward Detection of the
  Molecular Parity Violation in Chiral Ru(acac)3 and Os(acac)3. \emph{J.\
  Phys.\ Chem.\ Lett.} \textbf{2022}, \emph{13}, 10011--10017\relax
\mciteBstWouldAddEndPuncttrue
\mciteSetBstMidEndSepPunct{\mcitedefaultmidpunct}
{\mcitedefaultendpunct}{\mcitedefaultseppunct}\relax
\EndOfBibitem
\bibitem[Lehmann \latin{et~al.}(2013)Lehmann, Ram, Powis, and
  Janssen]{Lehmann:JCP139:234307}
Lehmann,~C.~S.; Ram,~N.~B.; Powis,~I.; Janssen,~M. H.~M. {Imaging photoelectron
  circular dichroism of chiral molecules by femtosecond multiphoton coincidence
  detection}. \emph{J.\ Chem.\ Phys.} \textbf{2013}, \emph{139}, 234307\relax
\mciteBstWouldAddEndPuncttrue
\mciteSetBstMidEndSepPunct{\mcitedefaultmidpunct}
{\mcitedefaultendpunct}{\mcitedefaultseppunct}\relax
\EndOfBibitem
\bibitem[Patterson \latin{et~al.}(2013)Patterson, Schnell, and
  Doyle]{Patterson:nature497:475}
Patterson,~D.; Schnell,~M.; Doyle,~J.~M. Enantiomer-specific detection of
  chiral molecules via microwave spectroscopy. \emph{Nature} \textbf{2013},
  \emph{497}, 475--477\relax
\mciteBstWouldAddEndPuncttrue
\mciteSetBstMidEndSepPunct{\mcitedefaultmidpunct}
{\mcitedefaultendpunct}{\mcitedefaultseppunct}\relax
\EndOfBibitem
\bibitem[Ayuso \latin{et~al.}(2022)Ayuso, Ordonez, and
  Smirnova]{Ayuso:PCCP24:26962}
Ayuso,~D.; Ordonez,~A.~F.; Smirnova,~O. Ultrafast chirality: the road to
  efficient chiral measurements. \emph{Phys.\ Chem.\ Chem.\ Phys.}
  \textbf{2022}, \emph{24}, 26962--26991\relax
\mciteBstWouldAddEndPuncttrue
\mciteSetBstMidEndSepPunct{\mcitedefaultmidpunct}
{\mcitedefaultendpunct}{\mcitedefaultseppunct}\relax
\EndOfBibitem
\bibitem[Janssen and Powis(2014)Janssen, and Powis]{Janssen:PCCP16:856}
Janssen,~M. H.~M.; Powis,~I. Detecting chirality in molecules by imaging
  photoelectron circular dichroism. \emph{Phys.\ Chem.\ Chem.\ Phys.}
  \textbf{2014}, \emph{16}, 856--871\relax
\mciteBstWouldAddEndPuncttrue
\mciteSetBstMidEndSepPunct{\mcitedefaultmidpunct}
{\mcitedefaultendpunct}{\mcitedefaultseppunct}\relax
\EndOfBibitem
\bibitem[Ganjitabar \latin{et~al.}(2018)Ganjitabar, Hadidi, Garcia, Nahon, and
  Powis]{Ganjitabar:jms353:11}
Ganjitabar,~H.; Hadidi,~R.; Garcia,~G.~A.; Nahon,~L.; Powis,~I.
  Vibrationally-resolved photoelectron spectroscopy and photoelectron circular
  dichroism of bicyclic monoterpene enantiomers. \emph{J.\ Mass.\ Spectrom.}
  \textbf{2018}, \emph{353}, 11--19\relax
\mciteBstWouldAddEndPuncttrue
\mciteSetBstMidEndSepPunct{\mcitedefaultmidpunct}
{\mcitedefaultendpunct}{\mcitedefaultseppunct}\relax
\EndOfBibitem
\bibitem[Chandler and Houston(1987)Chandler, and Houston]{Chandler:JCP87:1445}
Chandler,~D.~W.; Houston,~P.~L. Two-dimensional imaging of state-selected
  photodissociation products detected by multiphoton ionization. \emph{J.\
  Chem.\ Phys.} \textbf{1987}, \emph{87}, 1445--1447\relax
\mciteBstWouldAddEndPuncttrue
\mciteSetBstMidEndSepPunct{\mcitedefaultmidpunct}
{\mcitedefaultendpunct}{\mcitedefaultseppunct}\relax
\EndOfBibitem
\bibitem[Eppink and Parker(1997)Eppink, and Parker]{Eppink:RSI68:3477}
Eppink,~A. T. J.~B.; Parker,~D.~H. Velocity map imaging of ions and electrons
  using electrostatic lenses: Application in photoelectron and photofragment
  ion imaging of molecular oxygen. \emph{Rev.\ Sci.\ Instrum.} \textbf{1997},
  \emph{68}, 3477--3484\relax
\mciteBstWouldAddEndPuncttrue
\mciteSetBstMidEndSepPunct{\mcitedefaultmidpunct}
{\mcitedefaultendpunct}{\mcitedefaultseppunct}\relax
\EndOfBibitem
\bibitem[Ritchie(1976)]{Ritchie:PRA13:1411}
Ritchie,~B. Theory of the angular distribution of photoelectrons ejected from
  optically active molecules and molecular negative ions. \emph{Phys.\ Rev.\ A}
  \textbf{1976}, \emph{13}, 1411--1415\relax
\mciteBstWouldAddEndPuncttrue
\mciteSetBstMidEndSepPunct{\mcitedefaultmidpunct}
{\mcitedefaultendpunct}{\mcitedefaultseppunct}\relax
\EndOfBibitem
\bibitem[Cherepkov(1982)]{Cherepkov:CPL87:344}
Cherepkov,~N. Circular dichroism of molecules in the continuous absorption
  region. \emph{Chem.\ Phys.\ Lett.} \textbf{1982}, \emph{87}, 344--348\relax
\mciteBstWouldAddEndPuncttrue
\mciteSetBstMidEndSepPunct{\mcitedefaultmidpunct}
{\mcitedefaultendpunct}{\mcitedefaultseppunct}\relax
\EndOfBibitem
\bibitem[Powis(2000)]{Powis:JCP112:301}
Powis,~I. {Photoelectron circular dichroism of the randomly oriented chiral
  molecules glyceraldehyde and lactic acid}. \emph{J.\ Chem.\ Phys.}
  \textbf{2000}, \emph{112}, 301--310\relax
\mciteBstWouldAddEndPuncttrue
\mciteSetBstMidEndSepPunct{\mcitedefaultmidpunct}
{\mcitedefaultendpunct}{\mcitedefaultseppunct}\relax
\EndOfBibitem
\bibitem[Powis(2000)]{Powis:JPCA104:878}
Powis,~I. Photoelectron Spectroscopy and Circular Dichroism in Chiral
  Biomolecules: l-Alanine. \emph{J.\ Phys.\ Chem.\ A} \textbf{2000},
  \emph{104}, 878--882\relax
\mciteBstWouldAddEndPuncttrue
\mciteSetBstMidEndSepPunct{\mcitedefaultmidpunct}
{\mcitedefaultendpunct}{\mcitedefaultseppunct}\relax
\EndOfBibitem
\bibitem[B\"owering \latin{et~al.}(2001)B\"owering, Lischke, Schmidtke,
  M\"uller, Khalil, and Heinzmann]{Bowering:PRL86:1187}
B\"owering,~N.; Lischke,~T.; Schmidtke,~B.; M\"uller,~N.; Khalil,~T.;
  Heinzmann,~U. Asymmetry in Photoelectron Emission from Chiral Molecules
  Induced by Circularly Polarized Light. \emph{Phys.\ Rev.\ Lett.}
  \textbf{2001}, \emph{86}, 1187--1190\relax
\mciteBstWouldAddEndPuncttrue
\mciteSetBstMidEndSepPunct{\mcitedefaultmidpunct}
{\mcitedefaultendpunct}{\mcitedefaultseppunct}\relax
\EndOfBibitem
\bibitem[Garcia \latin{et~al.}(2003)Garcia, Nahon, Lebech, Houver, Dowek, and
  Powis]{Garcia:JCP119:8781}
Garcia,~G.~A.; Nahon,~L.; Lebech,~M.; Houver,~J.-C.; Dowek,~D.; Powis,~I.
  {Circular dichroism in the photoelectron angular distribution from randomly
  oriented enantiomers of camphor}. \emph{J.\ Chem.\ Phys.} \textbf{2003},
  \emph{119}, 8781--8784\relax
\mciteBstWouldAddEndPuncttrue
\mciteSetBstMidEndSepPunct{\mcitedefaultmidpunct}
{\mcitedefaultendpunct}{\mcitedefaultseppunct}\relax
\EndOfBibitem
\bibitem[Nahon \latin{et~al.}(2015)Nahon, Garcia, and
  Powis]{Nahon:JESRP204:322}
Nahon,~L.; Garcia,~G.~A.; Powis,~I. Valence shell one-photon photoelectron
  circular dichroism in chiral systems. \emph{J. Electron. Spectrosc. Relat.
  Phenom.} \textbf{2015}, \emph{204}, 322--334\relax
\mciteBstWouldAddEndPuncttrue
\mciteSetBstMidEndSepPunct{\mcitedefaultmidpunct}
{\mcitedefaultendpunct}{\mcitedefaultseppunct}\relax
\EndOfBibitem
\bibitem[Garcia \latin{et~al.}(2009)Garcia, Soldi-Lose, and
  Nahon]{Garcia:RSI80:023102}
Garcia,~G.~A.; Soldi-Lose,~H.; Nahon,~L. {A versatile electron-ion coincidence
  spectrometer for photoelectron momentum imaging and threshold spectroscopy on
  mass selected ions using synchrotron radiation}. \emph{Rev.\ Sci.\ Instrum.}
  \textbf{2009}, \emph{80}, 023102\relax
\mciteBstWouldAddEndPuncttrue
\mciteSetBstMidEndSepPunct{\mcitedefaultmidpunct}
{\mcitedefaultendpunct}{\mcitedefaultseppunct}\relax
\EndOfBibitem
\bibitem[Nahon \latin{et~al.}(2012)Nahon, {de Oliveira}, Garcia, Gil, Pilette,
  Marcouill{\'e}, Lagarde, and Polack]{Nahon:jsr19:508}
Nahon,~L.; {de Oliveira},~N.; Garcia,~G.~A.; Gil,~J.-F.; Pilette,~B.;
  Marcouill{\'e},~O.; Lagarde,~B.; Polack,~F. {{DESIRS}}: A State-of-the-Art
  {{VUV}} Beamline Featuring High Resolution and Variable Polarization for
  Spectroscopy and Dichroism at {{SOLEIL}}. \emph{J.\ Synchrotron\ Rad.}
  \textbf{2012}, \emph{19}, 508--520\relax
\mciteBstWouldAddEndPuncttrue
\mciteSetBstMidEndSepPunct{\mcitedefaultmidpunct}
{\mcitedefaultendpunct}{\mcitedefaultseppunct}\relax
\EndOfBibitem
\bibitem[Lux \latin{et~al.}(2012)Lux, Wollenhaupt, Bolze, Liang, Köhler,
  Sarpe, and Baumert]{Lux:ACIE51:5001}
Lux,~C.; Wollenhaupt,~M.; Bolze,~T.; Liang,~Q.; Köhler,~J.; Sarpe,~C.;
  Baumert,~T. Circular Dichroism in the Photoelectron Angular Distributions of
  Camphor and Fenchone from Multiphoton Ionization with Femtosecond Laser
  Pulses. \emph{Angew.\ Chem.\ Int.\ Ed.} \textbf{2012}, \emph{51},
  5001--5005\relax
\mciteBstWouldAddEndPuncttrue
\mciteSetBstMidEndSepPunct{\mcitedefaultmidpunct}
{\mcitedefaultendpunct}{\mcitedefaultseppunct}\relax
\EndOfBibitem
\bibitem[Vredenborg \latin{et~al.}(2008)Vredenborg, Roeterdink, and
  Janssen]{Vredenborg:RSI79:063108}
Vredenborg,~A.; Roeterdink,~W.~G.; Janssen,~M. H.~M. A photoelectron-photoion
  coincidence imaging apparatus for femtosecond time-resolved molecular
  dynamics with electron time-of-flight resolution of $\sigma$=18ps and energy
  resolution $\Delta${E}/{E}=3.5\%. \emph{Rev.\ Sci.\ Instrum.} \textbf{2008},
  \emph{79}, 063108\relax
\mciteBstWouldAddEndPuncttrue
\mciteSetBstMidEndSepPunct{\mcitedefaultmidpunct}
{\mcitedefaultendpunct}{\mcitedefaultseppunct}\relax
\EndOfBibitem
\bibitem[Fanood \latin{et~al.}(2015)Fanood, Ram, Lehmann, Powis, and
  Janssen]{Fanood:natcomm6:7511}
Fanood,~M. M.~R.; Ram,~N.~B.; Lehmann,~C.~S.; Powis,~I.; Janssen,~M. H.~M.
  Enantiomer-Specific Analysis of Multi-Component Mixtures by Correlated
  Electron Imaging\textendash Ion Mass Spectrometry. \emph{Nat. Commun.}
  \textbf{2015}, \emph{6}, 7511\relax
\mciteBstWouldAddEndPuncttrue
\mciteSetBstMidEndSepPunct{\mcitedefaultmidpunct}
{\mcitedefaultendpunct}{\mcitedefaultseppunct}\relax
\EndOfBibitem
\bibitem[Janssen \latin{et~al.}(2022)Janssen, Nahon, Smirnova, and
  Stolow]{Janssen:PCCP24:24611}
Janssen,~M.; Nahon,~L.; Smirnova,~O.; Stolow,~A. Fundamentals and applications
  of molecular photoelectron spectroscopy – Festschrift for Ivan Powis.
  \emph{Phys.\ Chem.\ Chem.\ Phys.} \textbf{2022}, \emph{24},
  24611--24613\relax
\mciteBstWouldAddEndPuncttrue
\mciteSetBstMidEndSepPunct{\mcitedefaultmidpunct}
{\mcitedefaultendpunct}{\mcitedefaultseppunct}\relax
\EndOfBibitem
\bibitem[Kastner \latin{et~al.}(2020)Kastner, Koumarianou, Glodic, Samartzis,
  Ladda, Ranecky, Ring, Vasudevan, Witte, Braun, Lee, Senftleben, Berger, Park,
  Schäfer, and Baumert]{Kastner:PCCP22:7404}
Kastner,~A. \latin{et~al.}  High-resolution resonance-enhanced multiphoton
  photoelectron circular dichroism. \emph{Phys.\ Chem.\ Chem.\ Phys.}
  \textbf{2020}, \emph{22}, 7404--7411\relax
\mciteBstWouldAddEndPuncttrue
\mciteSetBstMidEndSepPunct{\mcitedefaultmidpunct}
{\mcitedefaultendpunct}{\mcitedefaultseppunct}\relax
\EndOfBibitem
\bibitem[Ranecky \latin{et~al.}(2022)Ranecky, Park, Samartzis, Giannakidis,
  Schwarzer, Senftleben, Baumert, and Schäfer]{Ranecky:PCCP24:2758}
Ranecky,~S.~T.; Park,~G.~B.; Samartzis,~P.~C.; Giannakidis,~I.~C.;
  Schwarzer,~D.; Senftleben,~A.; Baumert,~T.; Schäfer,~T. Detecting chirality
  in mixtures using nanosecond photoelectron circular dichroism. \emph{Phys.\
  Chem.\ Chem.\ Phys.} \textbf{2022}, \emph{24}, 2758--2761\relax
\mciteBstWouldAddEndPuncttrue
\mciteSetBstMidEndSepPunct{\mcitedefaultmidpunct}
{\mcitedefaultendpunct}{\mcitedefaultseppunct}\relax
\EndOfBibitem
\bibitem[Ferr{\'e} \latin{et~al.}(2015)Ferr{\'e}, Handschin, Dumergue, Burgy,
  Comby, Descamps, Fabre, Garcia, G{\'e}neaux, Merceron, M{\'e}vel, Nahon,
  Petit, Pons, Staedter, Weber, Ruchon, Blanchet, and
  Mairesse]{Ferre:natphoton9:93}
Ferr{\'e},~A. \latin{et~al.}  A Table-Top Ultrashort Light Source in the
  Extreme Ultraviolet for Circular Dichroism Experiments. \emph{Nat. Photon.}
  \textbf{2015}, \emph{9}, 93--98\relax
\mciteBstWouldAddEndPuncttrue
\mciteSetBstMidEndSepPunct{\mcitedefaultmidpunct}
{\mcitedefaultendpunct}{\mcitedefaultseppunct}\relax
\EndOfBibitem
\bibitem[Sparling \latin{et~al.}(2023)Sparling, Crane, Ireland, Anderson,
  Ghafur, Greenwood, and Townsend]{Sparling:PCCP25:6009}
Sparling,~C.; Crane,~S.~W.; Ireland,~L.; Anderson,~R.; Ghafur,~O.;
  Greenwood,~J.~B.; Townsend,~D. Velocity-map imaging of photoelectron circular
  dichroism in non-volatile molecules using a laser-based desorption source.
  \emph{Phys.\ Chem.\ Chem.\ Phys.} \textbf{2023}, \emph{25}, 6009--6015\relax
\mciteBstWouldAddEndPuncttrue
\mciteSetBstMidEndSepPunct{\mcitedefaultmidpunct}
{\mcitedefaultendpunct}{\mcitedefaultseppunct}\relax
\EndOfBibitem
\bibitem[Comby \latin{et~al.}(2018)Comby, Bloch, Bond, Descamps, Miles, Petit,
  Rozen, Greenwood, Blanchet, and Mairesse]{Comby:natcomm9:5212}
Comby,~A.; Bloch,~E.; Bond,~C. M.~M.; Descamps,~D.; Miles,~J.; Petit,~S.;
  Rozen,~S.; Greenwood,~J.~B.; Blanchet,~V.; Mairesse,~Y. Real-Time
  Determination of Enantiomeric and Isomeric Content Using Photoelectron
  Elliptical Dichroism. \emph{Nat. Commun.} \textbf{2018}, \emph{9}, 5212\relax
\mciteBstWouldAddEndPuncttrue
\mciteSetBstMidEndSepPunct{\mcitedefaultmidpunct}
{\mcitedefaultendpunct}{\mcitedefaultseppunct}\relax
\EndOfBibitem
\bibitem[Comby \latin{et~al.}(2020)Comby, Bond, Bloch, Descamps, Fabre, Petit,
  Mairesse, Greenwood, and Blanchet]{Comby:chir32:1225}
Comby,~A.; Bond,~C.~M.; Bloch,~E.; Descamps,~D.; Fabre,~B.; Petit,~S.;
  Mairesse,~Y.; Greenwood,~J.~B.; Blanchet,~V. Using photoelectron elliptical
  dichroism (PEELD) to determine real-time variation of enantiomeric excess.
  \emph{Chirality} \textbf{2020}, \emph{32}, 1225--1233\relax
\mciteBstWouldAddEndPuncttrue
\mciteSetBstMidEndSepPunct{\mcitedefaultmidpunct}
{\mcitedefaultendpunct}{\mcitedefaultseppunct}\relax
\EndOfBibitem
\bibitem[Kr\"uger and Weitzel(2021)Kr\"uger, and Weitzel]{Kruger:acie60:17861}
Kr\"uger,~P.; Weitzel,~K.-M. Photoelectron Circular Dichroism in the
  Photodetachment of Amino Acid Anions. \emph{Angew.\ Chem.\ Int.\ Ed.}
  \textbf{2021}, \emph{60}, 17861--17865\relax
\mciteBstWouldAddEndPuncttrue
\mciteSetBstMidEndSepPunct{\mcitedefaultmidpunct}
{\mcitedefaultendpunct}{\mcitedefaultseppunct}\relax
\EndOfBibitem
\bibitem[Triptow \latin{et~al.}(2023)Triptow, Fielicke, Meijer, and
  Green]{Triptow:ACIE62:e202212020}
Triptow,~J.; Fielicke,~A.; Meijer,~G.; Green,~M. Imaging Photoelectron Circular
  Dichroism in the Detachment of Mass-Selected Chiral Anions. \emph{Angew.\
  Chem.\ Int.\ Ed.} \textbf{2023}, \emph{62}, e202212020\relax
\mciteBstWouldAddEndPuncttrue
\mciteSetBstMidEndSepPunct{\mcitedefaultmidpunct}
{\mcitedefaultendpunct}{\mcitedefaultseppunct}\relax
\EndOfBibitem
\bibitem[Kr\"uger \latin{et~al.}(2022)Kr\"uger, Both, Linne, Chirot, and
  Weitzel]{Kruger:jpcl13:6110}
Kr\"uger,~P.; Both,~J.~H.; Linne,~U.; Chirot,~F.; Weitzel,~K.-M. Photoelectron
  Circular Dichroism of Electrosprayed Gramicidin Anions. \emph{J.\ Phys.\
  Chem.\ Lett.} \textbf{2022}, \emph{13}, 6110--6116\relax
\mciteBstWouldAddEndPuncttrue
\mciteSetBstMidEndSepPunct{\mcitedefaultmidpunct}
{\mcitedefaultendpunct}{\mcitedefaultseppunct}\relax
\EndOfBibitem
\bibitem[Hickstein \latin{et~al.}(2016)Hickstein, Yurchak, Dhrubajyoti, Shih,
  and Gibson]{Hickstein:pyabel}
Hickstein,~D.~D.; Yurchak,~R.; Dhrubajyoti,~D.; Shih,~C.-Y.; Gibson,~S.~T.
  {PyAbel (v0.7): A Python Package for Abel Transforms}. 2016;
  \url{https://zenodo.org/record/47423}\relax
\mciteBstWouldAddEndPuncttrue
\mciteSetBstMidEndSepPunct{\mcitedefaultmidpunct}
{\mcitedefaultendpunct}{\mcitedefaultseppunct}\relax
\EndOfBibitem
\bibitem[Gibson \latin{et~al.}(2022)Gibson, Hickstein, Yurchak, Ryazanov, Das,
  and Shih]{rBasex:0.9.0}
Gibson,~S.; Hickstein,~D.; Yurchak,~R.; Ryazanov,~M.; Das,~D.; Shih,~G. rBasex
  method from the PyAbel package. 2022;
  {URL}:~\url{10.5281/zenodo.7438595}\relax
\mciteBstWouldAddEndPuncttrue
\mciteSetBstMidEndSepPunct{\mcitedefaultmidpunct}
{\mcitedefaultendpunct}{\mcitedefaultseppunct}\relax
\EndOfBibitem
\bibitem[Wallace(2017)]{NIST:webbook:MassSpec}
Wallace,~W.~E. In \emph{{NIST} Chemistry WebBook, {NIST} {S}tandard {R}eference
  {D}atabase {N}umber 69}; Linstrom,~P.~J., Mallard,~W.~G., Eds.; National
  Institute of Standards and Technology: Gaithersburg MD, 20899, 2017\relax
\mciteBstWouldAddEndPuncttrue
\mciteSetBstMidEndSepPunct{\mcitedefaultmidpunct}
{\mcitedefaultendpunct}{\mcitedefaultseppunct}\relax
\EndOfBibitem
\bibitem[Dauletyarov \latin{et~al.}(2023)Dauletyarov, Wang, and
  Horke]{Dauletyarov:JASMS34:1538}
Dauletyarov,~Y.; Wang,~S.; Horke,~D.~A. Vaporization of Intact Neutral
  Biomolecules Using Laser-Based Thermal Desorption. \emph{J.\ Am.\ Soc.\
  Mass.\ Spectrom.} \textbf{2023}, \emph{34}, 1538--1542\relax
\mciteBstWouldAddEndPuncttrue
\mciteSetBstMidEndSepPunct{\mcitedefaultmidpunct}
{\mcitedefaultendpunct}{\mcitedefaultseppunct}\relax
\EndOfBibitem
\bibitem[Mons \latin{et~al.}(2000)Mons, Piuzzi, Dimicoli, Zehnacker, and
  Lahmani]{Mons:PCCP2:5065}
Mons,~M.; Piuzzi,~F.; Dimicoli,~I.; Zehnacker,~A.; Lahmani,~F. Binding energy
  of hydrogen-bonded complexes of the chiral molecule 1-phenylethanol{,} as
  studied by 2C-R2PI: comparison between diastereoisomeric complexes with
  butan-2-ol and the singly hydrated complex. \emph{Phys.\ Chem.\ Chem.\ Phys.}
  \textbf{2000}, \emph{2}, 5065--5070\relax
\mciteBstWouldAddEndPuncttrue
\mciteSetBstMidEndSepPunct{\mcitedefaultmidpunct}
{\mcitedefaultendpunct}{\mcitedefaultseppunct}\relax
\EndOfBibitem
\bibitem[Cooper and Zare(2003)Cooper, and Zare]{Cooper:JCP48:942}
Cooper,~J.; Zare,~R.~N. {Angular Distribution of Photoelectrons}. \emph{J.\
  Chem.\ Phys.} \textbf{2003}, \emph{48}, 942--943\relax
\mciteBstWouldAddEndPuncttrue
\mciteSetBstMidEndSepPunct{\mcitedefaultmidpunct}
{\mcitedefaultendpunct}{\mcitedefaultseppunct}\relax
\EndOfBibitem
\bibitem[Greenwood and Williams(2023)Greenwood, and
  Williams]{Greenwood:PCCP25:16238}
Greenwood,~J.~B.; Williams,~I.~D. Investigation of photoelectron elliptical
  dichroism for chiral analysis. \emph{Phys.\ Chem.\ Chem.\ Phys.}
  \textbf{2023}, \emph{25}, 16238--16245\relax
\mciteBstWouldAddEndPuncttrue
\mciteSetBstMidEndSepPunct{\mcitedefaultmidpunct}
{\mcitedefaultendpunct}{\mcitedefaultseppunct}\relax
\EndOfBibitem
\bibitem[Beaulieu \latin{et~al.}(2016)Beaulieu, Ferré, Géneaux, Canonge,
  Descamps, Fabre, Fedorov, Légaré, Petit, Ruchon, Blanchet, Mairesse, and
  Pons]{Beaulieu:njp18:102002}
Beaulieu,~S.; Ferré,~A.; Géneaux,~R.; Canonge,~R.; Descamps,~D.; Fabre,~B.;
  Fedorov,~N.; Légaré,~F.; Petit,~S.; Ruchon,~T.; Blanchet,~V.; Mairesse,~Y.;
  Pons,~B. Universality of photoelectron circular dichroism in the
  photoionization of chiral molecules. \emph{New J.\ Phys.} \textbf{2016},
  \emph{18}, 102002\relax
\mciteBstWouldAddEndPuncttrue
\mciteSetBstMidEndSepPunct{\mcitedefaultmidpunct}
{\mcitedefaultendpunct}{\mcitedefaultseppunct}\relax
\EndOfBibitem
\end{mcitethebibliography}
\bibliographystyle{achemso}

\end{document}